Oxygen depletion hypothesis remains controversial: a mathematical model of oxygen depletion during FLASH radiation

Ankang Hu[1,2], Rui Qiu[1,2,*], Zhen Wu[1,3], Chunyan Li[1,3], Hui Zhang[1,2], Junli Li[1,2]

[1.] Department of Engineering Physics, Tsinghua University, Beijing, China
[2.] Key Laboratory of Particle & Radiation Imaging, Tsinghua University, Ministry of Education, Beijing, China
[3.] Nuctech Company Limited, Beijing, China

Corresponding author:
Rui Qiu: qiurui@mail.tsinghua.edu.cn

Background: Experiments have reported low normal tissue toxicities during FLASH radiation, but the mechanism has not been elaborated. Several hypotheses have been proposed to explain the mechanism. The oxygen depletion hypothesis has been introduced and mostly studied qualitatively.
Methods: We present a computational model to describe the time-dependent change of oxygen concentration in the tissue. The kinetic equation of the model is solved numerically using the finite difference method. The model is used to analyze the FLASH effect with the oxygen depletion hypothesis, and the brain tissue is chosen as an example.
Results: The oxygen distribution is determined by the oxygen consumption rate of the tissue and the distance between capillaries. The change of oxygen concentration with time after radiation has been found to follow a negative exponential function, and the time constant is determined by the distance between capillaries. When the dose rate is high enough, the same dose results in the same change of oxygen concentration regardless of dose rate. The analysis of FLASH effect in the brain tissue based on this model does not support the explanation of the oxygen depletion hypothesis.
Conclusions: The oxygen depletion hypothesis remains controversial because oxygen in most normal tissues cannot be depleted by FLASH radiation according to the mathematical analysis with this model and experiments on the expression and distribution of the hypoxia-inducible factors.

## Introduction

Some animal experiments have reported unexpectedly low normal tissue toxicities during ultra-high dose rate radiation, which is called FLASH effect [1,2,3]. Because of the great potential clinical benefit resulted from the low damage to normal tissue, FLASH radiotherapy is attracting great attention in the radiation oncology community[4]. However, a mechanism explaining the FLASH effect has not yet been demonstrated[3,4]. Targeted experiments directed by the explanation can be performed to elucidate the mechanism of FLASH effect and find more situations showing FLASH effect. Oxygen depletion hypothesis has been proposed to explain the FLASH effect[5,6,7,8]. The hypothesis assumes that cells irradiated by ultra-high dose rate radiation become hypoxic so that they show radiation resistance. Several groups have tried to explain the FLASH effect according to the oxygen depletion hypothesis[5,6,7,8]. The oxygen depletion hypothesis has been used to predict the conditions that may lead to FLASH effect such as the pulse width, the interval between two pulses, the total dose of the radiation beam, and the original oxygen tension in the tissue[7]. However, the oxygen depletion hypothesis was put forward qualitatively[5,6,8]. There is no sufficient quantitative analysis to confirm this hypothesis. Considering that oxygen diffusion and metabolism are dynamic processes, a kinetic equation describing these two processes is required. Oxygen partial pressure in the tissue is strongly affected by the chemical reaction rate of free radicals not just the yield and reaction types of free radicals. Qualitative analysis can hardly give a reliable conclusion. Therefore, detailed analysis is required to estimate the influence of the radiation conditions, and the hypothesis should be investigated carefully.

Very recently Pratx and Kapp have proposed a computational model to calculate the distribution of oxygen surrounding a single capillary and to predict the biological effect based on the oxygen enhancement model[7]. However, we find that there are some limitations in Pratx and Kapp's model. First, a boundary condition that is necessary for solving the equation and reflecting the microvessel density in tissue is missed in the model. Moreover, a constant oxygen consumption rate under any oxygen concentration in their model is inconsistent with the fact. A new model is proposed in this work to overcome these limitations. Our new model gives a more rational description of the time-dependent change of oxygen distribution in the tissue. In this study, the brain, in which FLASH effect has been observed[2], is chosen as an example. The time-dependent change of oxygen distribution and the biological effect are obtained with this new model. The relationship between the time-dependent change of oxygen distribution after irradiation and some biological parameters is also explored. It turns out that completely different results and an opposite conclusion compared with the work by Pratx and Kapp are obtained from our new model. The oxygen depletion hypothesis may not be the truth of the FLASH effect according to the results of this work.

**Theory**

As mentioned above, recently Pratx and Kapp have established a computational model to calculate the distribution of oxygen surrounding a single capillary and to predict biological effect based on the oxygen enhancement model[7]. Although the model in this work is similar to their work, a complete illustration of our model is provided as following because there are large differences in boundary conditions and oxygen consumption. An opposite conclusion is drawn due to these differences.

**Oxygen consumption of tissue**

The reaction-diffusion model is commonly used to calculate oxygen distribution in the tissue. The model is depended on the oxygen consumption rate (OCR) of the tissue[9]. There is no exact curve to describe OCR versus oxygen concentration. We choose a negative exponential function as the OCR function, which is easy to calculate its derivative. The OCR of the tissue is given by

$$OCR(p) = OCR_{max}(1 - \exp(-\lambda p)) \qquad (1)$$

where $p$ is the oxygen concentration, $OCR_{max}$ is the maximum oxygen consumption rate and $\lambda$ is a constant.

OCR of the tissue tends to be zero when the oxygen tension tends to zero and shows a tendency of saturation when the oxygen concentration increases[10]. When the oxygen concentration is about 1.4 μmol/L(1 mmHg), OCR is observed at the half level of $OCR_{max}$ typically[11]. Any curves showing a similar form can be used as the OCR function. We compared our OCR function with the hyperbolic curve, the differences are within 10% for oxygen concentration ranging from 10 to 200 μmol/L. Some models regard the *OCR* as a constant[7,11], but the constant *OCR* may lead to a negative value of oxygen tension in the reaction-diffusion model. Therefore, the negative exponential function, formula(1), is used to describe the relationship between OCR and the oxygen concentration in our model.

**Simplified tissue-capillary model**

The structure of the capillary system is quite complicated. There exist some problems to calculate the distribution of oxygen in the tissue accurately, such as establishing the detailed model of capillaries in the tissue and solving the kinetic equation of oxygen in such complex geometry. A simplified model is required to calculate the oxygen distribution in the tissue approximately[12]. We use a Krogh geometrical model, which is widely used model to describe the capillary-tissue system[13]. The model regards a capillary and the surrounding tissue as two coaxial cylinders. The inner cylinder represents the capillary and the outer cylinder represents the tissue around the capillary. The geometrical model is shown in figure 1.

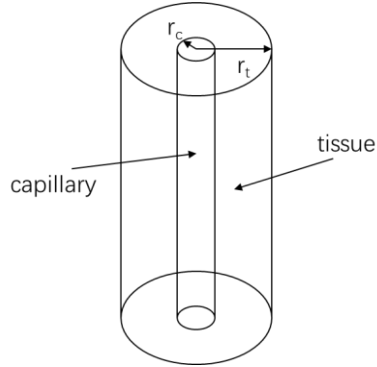

Figure 1. Diagram of Krogh cylinder

We further assume that the tissue around the capillary is homogeneous, which means they have the same composition and oxygen consumption rate. Then the oxygen diffusion and consumption in the tissue can be described by

$$\frac{\partial p}{\partial t} = D\nabla^2 p - OCR(p) \qquad (2)$$

where $p$ is the molar concentration of $O_2$ in the tissue, $D$ is the diffusion coefficient of $O_2$, $OCR(p)$ is the function of oxygen consumption rate related to the $p$. The steady-state transport of $O_2$ is governed by

$$D\nabla^2 p - OCR(p) = 0 \qquad (3)$$

Considering that the radius of the outer cylinder(<100μm) is far less than the length of the capillary(~1mm), we assume that the axial diffusion of $O_2$ in the tissue is negligible compared to the radial diffusion. Oxygen distribution in the tissue can be described in the cross-section plane with polar coordinates. The steady-state equation is simplified to

$$D\left(\frac{\partial^2 p}{\partial r^2} + \frac{1}{r}\frac{\partial p}{\partial r}\right) - OCR(p) = 0 \qquad (4)$$

Equation (4) is subject to the following boundary conditions

$$p = p_w, r = r_c$$
$$\frac{\partial p}{\partial r} = 0, r = r_t \tag{5}$$

where $p_w$ is the molar concentration of $O_2$ at the capillary-tissue interface, $r_c$ is the radius of the capillary, $r_t$ is obtained from the value of microvessel density. The value of $r_t$ is approximately equal to the half distance between two adjacent capillaries. The fluxes of oxygen from these two adjacent capillaries are equal at the position $r = r_t$, so that the first derivative of $p$ versus $r$ is equal to zero. One of the main differences between the model in this work and Pratx and Kapp. is the boundary condition[7].

A finite difference method was adopted to solve the equation (3) with the boundary conditions. The steady-state oxygen distribution in the tissue can be obtained from the solution.

**Time-dependent change of oxygen distribution during FLASH**

The steady-state equation gives the oxygen distribution in the tissue without radiation. To consider the change of oxygen distribution in the tissue receiving FLASH radiation, we add a reaction term to the steady-state equation. Radiation generates reducing radicals, such as the •H and $e^-_{aq}$, which can react with oxygen to yield superoxide ($O_2$•-) and its protonated form ($HO_2$•). We assume that the radicals yielded by low-LET radiation are homogeneous in the tissue[5]. The time-dependent change of oxygen distribution is governed by

$$\frac{\partial p}{\partial t} = D\left(\frac{\partial^2 p}{\partial r^2} + \frac{1}{r}\frac{\partial p}{\partial r}\right) - OCR(p) - kpL(t) \tag{6}$$

where $k$ is the reaction rate constant of oxygen-radical reaction, $L(t)$ is the molar concentration of the reducing radicals. The molar concentration of the reducing radicals $L(t)$ is described by

$$\frac{\partial L(t)}{\partial t} = R(t) - kpL(t) \tag{7}$$

where $R(t)$ is the yielding rate of reducing radicals, which is directly proportional to the dose rate. We assume that the oxygen in capillaries can hardly be affected by radiation because the hemoglobin carries a large amount of oxygen compared with the dissolved oxygen in the tissue[13]. We solve equation (6) and (7) with the same boundary conditions used in the steady-state equation. Moreover, the steady-state oxygen distribution is used as the initial value, and the initial value of the reducing radicals is set to be zero.

The Crank-Nicolson method is used to solve the time-dependent equation of oxygen distribution. Solutions of the time-dependent equation reflect the effect of radiation on oxygen distribution. We can quantitatively analyze the oxygen depletion hypothesis using these solutions.

To explore the relationship between the time-dependent change of oxygen distribution and some biological parameters such as microvessel density and oxygen consumption rate, we set different values of $r_t$ and $OCR_{max}$.

**Biological effect predicted by oxygen enhancement model**

The radiosensitivity is significantly affected by the oxygen concentration for low-LET radiation. Several models have been proposed to fit experimental cell survival data according to the oxygen level. We choose a classical model to predict the biological effect of FLASH radiation. Oxygen enhancement ratio (OER) is defined to represent the radiosensitivity of the tissues on different

oxygen levels. In the mentioned classical model[14], OER is given by

$$OER = \frac{mp + K}{p + K} \quad (8)$$

Where $p$ is the molar concentration of oxygen, μmol/L; $m$ and $K$ are constant parameters, 2.9 and 7.2 μmol/L, respectively.

We can calculate the OER based on the model and the solution of the time-dependent equation to estimate the biological effect of FLASH according to the oxygen depletion hypothesis.

**Parameters of the model**

Studies on brain tissue have measured the data of biological parameters of the brain such as the OCR and the microvessel density of brain tissue[15,16]. We choose the brain tissue as an example to analyze oxygen consumption during FLASH radiation. The model is solved with typical values of microvessel density, $OCR_{max}$, and oxygen concentration at the venous end of the capillary in the brain tissue. If we assume that capillaries are arranged in parallel, the boundary condition $r_t$ can be calculated by length density of capillary. The microvessel density is converted to the boundary condition $r_t$ by

$$r_t = \frac{1}{f\sqrt{L_v}} \quad (9)$$

where $L_v$ is the length density of capillary[15], $f$ is a factor related to the form of arrangement. If the capillaries are arranged in the form of square lattice in the cross-section plane, $f$ is equal to 2.0. If the capillaries are arranged in the form of a hexagonal lattice, $f$ is equal to 2.278. We choose 2.0 as the value of $f$, which leads to more zones at a low oxygen level. If the result of the model using this parameter showed no hypoxic zone, the model using a larger value of $f$ would not give the result showing a hypoxic zone. A typical value of $L_v$ in the cerebral cortex, 500 mm$^{-2}$, is chosen to calculate the boundary condition.

The constant dose rate during the pulse is assumed for its simplicity. The combined yield of solvated electrons and hydrogen radicals, $G$, is about 0.5 μmol/h/Gy[7]. Yielding rate of radicals, $R(t)$, is given by

$$R(t) = G \cdot Rate(t) \quad (10)$$

Where $Rate(t)$ is the dose rate, Gy/ms. Parameters of the model are listed in table 1.

Table 1 Parameters of the model for the brain tissue [7]

| Parameter | Value |
| --- | --- |
| $OCR_{max}$ | $20 \times 10^{-3}$ μmol/L/ms[16] |
| $\lambda$ | 0.485 L/μmol[11] |
| $D$ | 2 μm$^2$/ms[7] |
| $p_w$ | 57 μmol/L[11] |
| $r_c$ | 3 μm[7] |
| $r_t$ | 28 μm[15] |
| $f$ | 2.0 |
| $k$ | $1 \times 10^9$ L/mol[5] |
| $G$ | 0.5 μmol/h/Gy[7] |
| $m$ | 2.9[13] |

| $K$ | 7.2 μmol/L[14] |

With these typical parameters, we analyzed the time-dependent change of oxygen concentration and the oxygen depletion hypothesis on the brain tissue.

## Results and discussion

### Oxygen distribution in the tissue

We solved the steady-state equation of oxygen distribution with different values of $r_t$ and $OCR_{max}$ to evaluate the relationship between oxygen distribution and biological parameters. The results are shown in figure 2.

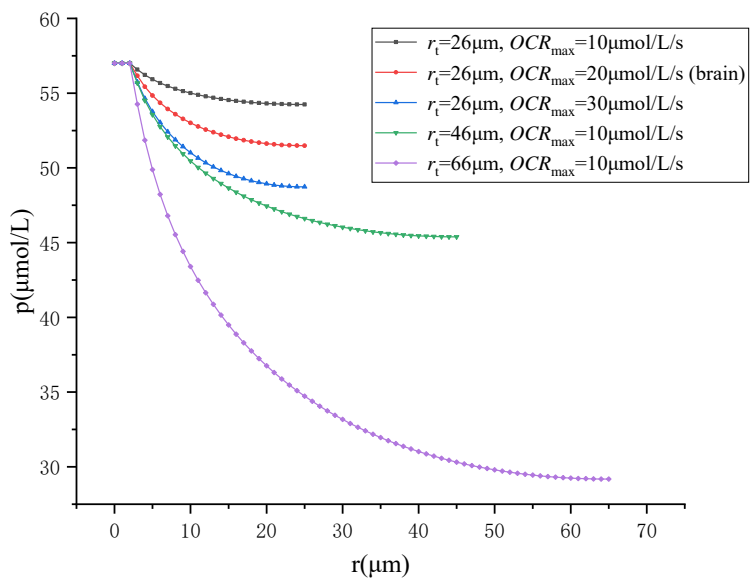

Figure 2. Oxygen distribution for different $r_t$ and $OCR_{max}$

Results show that oxygen distribution is significantly affected by oxygen consumption rate and the boundary condition. For the brain tissue, the value of oxygen concentration is higher than the level (< 20 μmol/L) at which the radiation-resistant phenomenon can be observed.

### Characteristic of time-dependent change of oxygen distribution

The time-dependent equation was solved with different parameters to evaluate the influence of biological parameters. For each situation, the tissue receives a pulse with a total dose of 20 Gy and a pulse width of 50 ms. The time-dependent change of oxygen at the position $r = 18$μm is shown in figure 3.

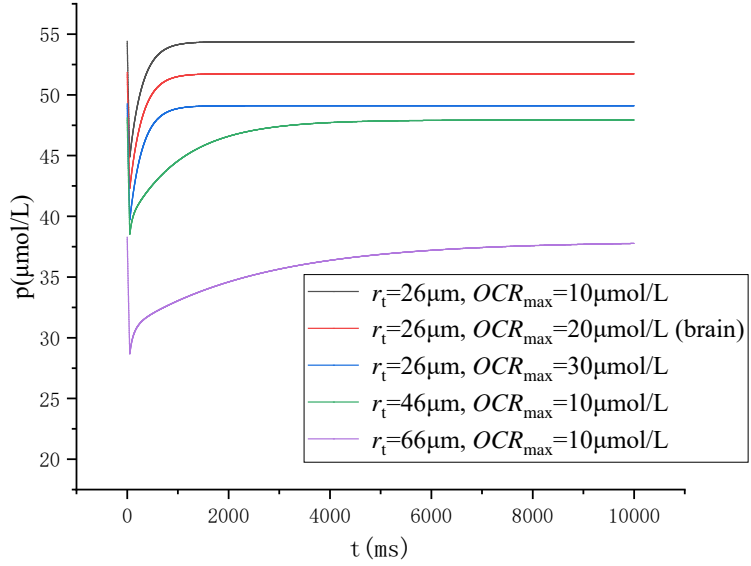

Figure 3. Time-dependent change of oxygen concentration for different biological parameters

We found that the change of oxygen concentration is approximately a negative exponential curve after the pulse, which can be described by

$$p(t) = p_s - \Delta p \cdot e^{-t/\tau} \qquad (11)$$

where $p_s$ is the steady-state oxygen concentration, $\Delta p$ is the oxygen consumed by radiation pulse and $\tau$ is the time constant of oxygen concentration recovering. Besides, the values of time constant at the positions $r$=23, 43 and 63μm for $r_t$=66μm are calculated. Table 2 shows the values of $\tau$ at different positions for different biological parameters.

Table 2 Values of time constant at different positions for different biological parameters

| $r_t$(μm) | $OCR_{max}$(μmol/L/s) | $r$(μm) | $\tau$(ms) |
| --- | --- | --- | --- |
| 26 | 10 | 18 | 255 |
| 26 | 20 | 18 | 255 |
| 26 | 30 | 18 | 255 |
| 46 | 10 | 18 | 1072 |
| 66 | 10 | 18 | 2479 |
| 66 | 10 | 23 | 2502 |
| 66 | 10 | 43 | 2607 |
| 66 | 10 | 63 | 2660 |

We further found that $\tau$ is independent on $OCR_{max}$ but related to $r_t$ and the position $r$. The distance to the capillary $r$ affects $\tau$ slightly at the position far away from the capillary, compared to the influence of the $r_t$. It indicates that the rule of time-dependent change of oxygen concentration is closely related to $r_t$, which is determined by the value of microvessel density. Moreover, theoretical analysis based on formula (11) indicates that oxygen consumed by the radiation pulse is no longer related to the width of the pulse, if the pulse width is much less than $\tau$. It means that the same dose results in the same change of oxygen concentration when the dose rate is high enough. We set a situation as an example, in which the tissue receives radiation pulses with the same dose but different pulse width. The time-dependent change of oxygen concentration is shown in figure 4. It can be seen in the figure that the minimum value of oxygen concentration is almost independent of the pulse width. The difference in the minimum value of oxygen concentration in figure 4 is less than 1%.

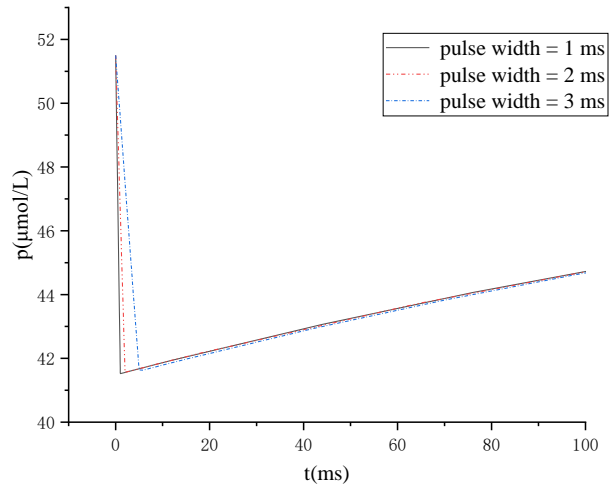

Figure 4. Time-dependent change of oxygen concentration for the same dose with different pulse width (pulse width = 1, 2, 5 ms, $r_t$ = 26 μm, $OCR_{max}$ = 20 μmol/L/s, $r$ = 18 μm)

For multi-pulse and wide pulse situations, formula (11) can be regarded as the impulse response function of a system so that the time-dependent curve can be obtained by convolution, which is widely used in the analysis of a signal system.

**Biological effect predicted by oxygen enhancement model**

We can calculate the value of OER of each position when receiving FLASH radiation and compare it with the OER under steady-state oxygen concentration. If the pulse width is several milliseconds, the OER is calculated by[7]

$$OER = \frac{\int_0^T OER(p(t)) \cdot Rate(t) dt}{\int_0^T Rate(t) dt} \tag{12}$$

where $T$ is the width of the pulse. We calculated the OER of the brain tissue in the previously calculated FLASH pulse (20 Gy, 50 ms). OER values at different positions of the brain tissue are compared with the steady-state OER in figure 5.

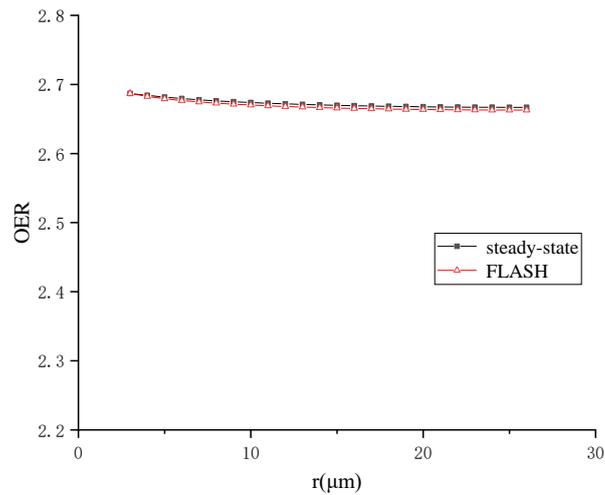

Figure 5. OER of steady-state oxygen and FLASH radiation at different positions of the brain tissue under oxygen depletion hypothesis

The differences of OER values are within 1% between FLASH radiation and conventional radiation under the steady-state oxygen concentration. There is no obvious change of OER for the brain tissue receiving FLASH radiation. The reason could be explained as follows. The steady-state oxygen concentration is higher than 50 μmol/L at every position, and the 20 Gy radiation pulse just consumes 10 μmol/L oxygen. The values of oxygen concentration in both steady-state and FLASH radiation are much higher than the interval(< 20 μmol/L), in which radiation resistance caused by hypoxia is significant.

**Oxygen depletion hypothesis remains controversial**

The oxygen depletion hypothesis states that FLASH radiation consumes oxygen in the tissue so that the tissue shows radiation resistance because of hypoxia. The original oxygen concentration and oxygen consumption by the radiation are two important factors. According to the data of experiments, OER changes significantly when the oxygen concentration is lower than 20 μmol/L (figure 6)[14]. Considering that a typical dose of FLASH is about 20 Gy (consuming 10 μmol/L oxygen), the original oxygen concentration should be less than 25 μmol/L (17.5 mmHg) so that significant change of radiosensitivity determined by oxygen can be observed[14].

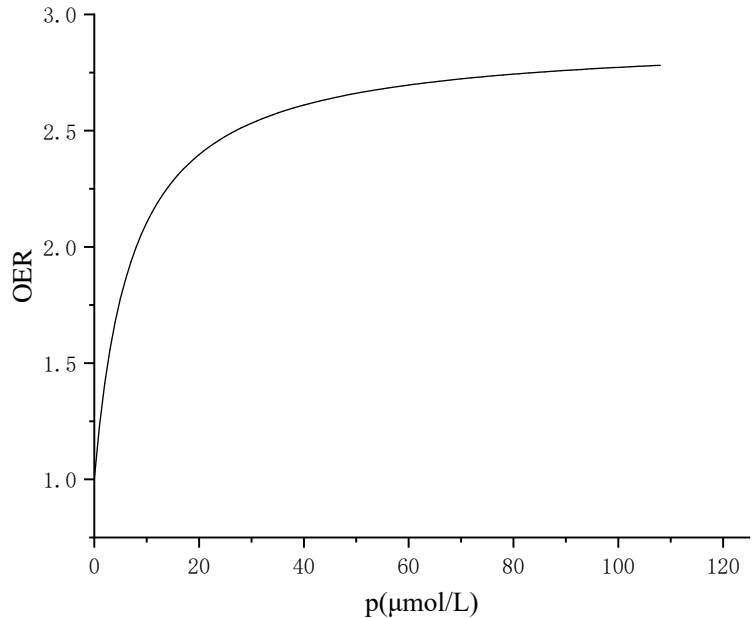

Figure 6. Curve of OER vs oxygen concentration

Our results show that the values of original oxygen concentration at all positions in the brain are much higher than 25 μmol/L, but FLASH effect has been observed in the brain tissue.

The divergence between our model and the existing model is the difference in boundary conditions[7]. Their model was solved with just one boundary condition (the oxygen concentration at the capillary). However, the unique solution of the second-order differential equation cannot be determined by only one boundary condition. To get a unique solution, we add a boundary condition, which is determined by microvessel density. We think that the capillary is not isolated, and our boundary conditions consider the effect of surrounding capillaries. The microvessel density and the oxygen consumption rate of tissues vary greatly, both of which strongly affect the oxygen distribution. Our model can be used to analyze the tissue with different values of microvessel density by introducing the additional boundary condition. Recent experiments have reported FLASH effect in lung and brain. There is no significant evidence showing that a large zone of hypoxic tissue can be observed in these organs. Results of experiments on expression and distribution of the hypoxia-inducible factors (HIF) also show that most zones in the normal tissue are not hypoxic[17]. The quantitative analysis of the brain tissue in this does not support the oxygen depletion hypothesis, either.

**Limitations of this work**

It should be noted that there are still some limitations in our new model. The main limitation is brought by the Krogh cylinder model. The simple Krogh cylinder model describing the capillary-tissue system cannot consider the complicated geometry of the real capillary system. Some experiments have reported that there are hypoxic microenvironments in neural stem cell niches (oxygen concentration in the stem cell niches between 10-80 μmol/L)[18], which cannot be explained by the Krogh cylinder model. The uncertainty of microvessel density and oxygen consumption and the assumption of uniform distribution of radicals may lead to the difference of results. However, the oxygen concentration in the brain tissue calculated by our model is larger than 50 μmol/L at every position. These values are far beyond the range in which OER changes obviously (< 20 μmol/L). Therefore these limitations can hardly affect the conclusion.

Analysis of other tissues is not performed due to the lack of data on biological parameters. The model can be used to analyze the time-dependent change of oxygen concentration in other tissues as long as the biological parameters are available.

**New Hypothesis**

FLASH effect has been observed in organs that are not hypoxic such as lung and brain. Experiments with lung fibroblasts in vitro have reported changes of senescence and cytokines (TGF-β) but have not observed significant change of cell survival after FLASH radiation[19]. These results suggest that the main reason for FLASH effect observed in vivo may be independent of the oxygen. The experiment on mice has reported that FLASH prevents activation of the TGF-β/SMAD cascade, and Favaudon et al have suggested a hypothesis considering the role of poly(ADP-ribose) polymerases (PARPs)[1]. PARPs work as the DNA damage sensor and a link between DNA damage and TGF-β/SMAD cascade[20]. We further hypothesize that the FLASH effect is induced by the saturation of the DNA damage sensor (DNA damage sensor saturation hypothesis), for example, PARP is one kind of DNA damage sensor[20,21]. FLASH radiation causes a large amount of DNA damage in a short time so that just a part of damage activates most of the sensors, which leads to incomplete DNA damage signal transduction. This incomplete signal transduction results in lower level response compared to that induced by the conventional radiation with the same dose. For example, FLASH radiation does not trigger the activation of the TGF-β/SMAD cascade like conventional radiation[1,22]. A large amount of DNA damage in a short time induced by FLASH radiation makes the damage sensor saturated so that the tissues respond as if they receive low dose radiation. Quantitative experiments on the response of FLASH radiation are required to test this hypothesis.

## Conclusion

A new model is established to describe the distribution and the kinetic of oxygen in the tissue on the scale of microvessel. This new model overcomes the limitations of ignoring the influence of surrounding capillaries and regarding the oxygen consumption rate of tissue as a constant in previous models. This model is used to analyze FLASH effect in brain tissue combining with the oxygen enhancement model.

The study on the oxygen concentration in the tissue using this model gives some conclusions on the time-dependent change of distribution: (1)the steady-state oxygen distribution is determined by the values of oxygen consumption rate and distance between capillaries. (2)The oxygen concentration recovers to the steady-state with a negative exponential format after radiation. (3)The time constant of the exponential format is only determined by the distance between capillaries. (4)The same dose results in the same change range of oxygen concentration regardless of dose rate, if the pulse width is much less than the time constant.

The analysis for brain tissue using this model does not support the explanation of the oxygen depletion hypothesis to the observed FLASH effect. The oxygen concentration in brain tissue calculated by this model is supported by the measurement of HIF. Therefore we think the oxygen depletion hypothesis remaining controversial. Based on the experiments in vivo and in vitro, the DNA damage sensor saturation hypothesis is promising to explain the FLASH effect and suggested for the future studies.